\documentclass[pre, preprint,groupedaddress, showkeys, showpacs]{revtex4}

\usepackage{graphicx}
\begin{document}


\title{Brownian pump in nonlinear diffusive media}


\author{Bao-quan  Ai$^{1}$}\email[Email: ]{aibq@hotmail.com}
\author{Liang-gang Liu$^{2}$}

\affiliation{$^{1}$Institute for Condensed Matter Physics, School of
Physics and Telecommunication Engineering, South China
Normal University, 510006 Guangzhou, China\\
$^{2}$ Faculty of Information Technology , Macau University of
Science and Technology, Macao}


\date{\today}
\begin{abstract}
\indent A Brownian pump in nonlinear diffusive media is investigated
in the presence of an unbiased external force. The pumping system is
embedded in a finite region and bounded by two particle reservoirs.
In the adiabatic limit, we obtain the analytical expressions of the
current and the pumping capacity as a function of temperature for
normal diffusion, subdiffusion and superdiffusion. It is found that
important anomalies are detected in comparison to the normal
diffusion case. The superdiffusive regime, compared with the normal
one, exhibits an opposite current for low temperatures. In
subdiffusive regime, the current may become forbidden for low
temperatures and negative for high temperatures.

\end{abstract}




\maketitle


\section{Introduction}

\indent Pumping is an active nonequilibrium transport process in
which fluctuations play a very important role. Some pumps have
already been investigated in the literature\cite{1,2,3,4,5,6}.
Kosztin and Schulten \cite{1} studied the fluctuation-driven
molecular transport through an asymmetric potential pump and three
transport mechanisms: driven by potential gradient, by an external
periodic force and by nonequilibrium fluctuations \cite{2}.
Moskalets and Buttiker \cite{3} studied heat fluxes in adiabatic
quantum pumps and developed a approach to the kinetics for an
arbitrary relation of pump frequency and temperature. Arrachea and
coworkers\cite{4} investigated the heat transport and the power
developed the local driving fields on an quantum system coupled
macroscopic reservoirs. They identified two generic interesting
mechanisms: directed heat transport between reservoirs induced by
the ac potentials and at slow driving, two oscillating out of phase
forces perform work against each other. Nonadiabatic electron heat
pump was investigated by Rey and coworkers \cite{5}. They presented
a mechanism for extracting heat metallic conductors based on the
energy-selective transmission of electrons through a spatially
asymmetric resonant structure subject to ac driving.

\indent Recently, Sancho and Gomez-marin \cite{6} presented a model
for a Brownian pump powered by a flashing ratchet mechanism. The
pumping device was embedded in a finite region and bounded by
particle reservoirs. Their emphasis is on finding what concentration
gradient the pump can maintain. All of the above examples have been
formulated within a standard Brownian framework, for which diffusion
properties are normal, that is, with the mean quadratic displacement
growing linearly with time $t$. However, it is well know that there
are media signaled by a $t^{\mu}$ growth of the squared dispersion
with $\mu\neq 1$. An important class is given by the " porous medium
" equation \cite{7,8,9,10}, which, in the one-dimensional problem
and in the absence of external forces, can cast in the form
\begin{equation}\label{}
    \partial_{t}\rho=D\mu\partial_{x}(\rho^{\mu-1}\partial_{x}\rho),
\end{equation}
where $\rho$ is the density of the diffusing substance, $x$ is a
dimensionless coordinate representing a bond length, angle, or any
other chemical or physical state variable, $t$ is the dimensionless
time, and $D$ is diffusive constant.  The mean quadratic deviation
follows the law $\langle x(t)^{2}\rangle\propto t^\frac{2}{1+\mu}$.
The transport is subdiffusive for $\mu>1$, normal for $\mu=1$ and
superdiffusive for $0<\mu<1$. Many physical systems are
well-described by this class of processes: percolation of gases in
porous media \cite{11}, dispersion of biological populations
\cite{12}, grain segregation \cite{13}, fluxes in plasma \cite{14}
and nonextensive statistics \cite{15}.

\indent The present work is extend to the study of the Brownian pump
to the case of the anomalous diffusion. We emphasize on finding how
particles can be pumped through a cell membrane from a particle
reservoir at low concentration to one at the same or higher
concentration in a nonlinear diffusive media.

\section {General analysis}
\indent We consider a Brownian pump which is located in a finite
region and transports particles across the barrier against the
concentration gradients by using an unbiased external force. Here,
we model the device as a pumping system in which overdamped Brownian
particles moving in a asymmetric finite potential in the presence of
an unbiased external force. The potential is embedded in a finite
region $[0, L]$,
\begin{figure}[htbp]
 \begin{center}\includegraphics[width=10cm,height=4cm]{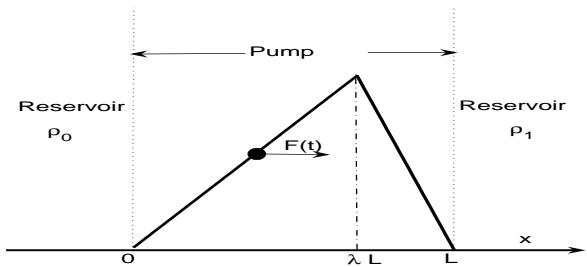}
\caption{Scheme of a Brownian pump: A spatially asymmetric potential
$U_{0}(x)$(defined in Eq. (2)) is embedded in a finite region and
bounded by two particle reservoirs of concentrations $\rho_{0}$ and
$\rho_{1}$. $L$ and $\lambda$ is the length and the asymmetric
parameter of the potential, respectively.  The particles are driven
by an unbiased external force $F(t)$, defined by Eq. (3).}\label{1}
\end{center}
\end{figure}

\begin{equation}\label{}
 U_{0}(x)=\left\{
\begin{array}{ll}
   Q\frac{x}{\lambda L},& \hbox{$0 \leq
x<\lambda L$};\\
   Q\frac{L-x}{(1-\lambda)L} ,&\hbox{$\lambda L\leq x \leq L$},\\
\end{array}
\right.
\end{equation}
where $Q$ is amplitude of the potential and $\lambda$ is its
asymmetric parameter. $F(t)$ is an unbiased external force and
satisfies
\begin{equation}\label{}
 F(t)=\left\{
\begin{array}{ll}
   F_{0},& \hbox{$n\tau\leq
t<n\tau+\frac{1}{2}\tau$};\\
   -F_{0} ,&\hbox{$n\tau+\frac{1}{2}\tau<t\leq
      (n+1)\tau$},\\
\end{array}
\right.
\end{equation}
where $\tau$ is the period of the unbiased force and $F_{0}$ is its
magnitude.

\indent The correlated anomalous diffusion can be described through
the following nonlinear Fokker-Planck equation \cite{8,9,10}:

\begin{equation}\label{}
    \frac{\partial}{\partial t}\rho(x,t)=\frac{\partial}{\partial
    x}[U^{'}(x,t)\rho(x,t)]+D\frac{\partial^{2}}{\partial
    x^{2}}\rho^{\mu}(x,t)=-\frac{\partial j(x,t)}{\partial x},
\end{equation}

\begin{equation}\label{}
    U(x,t)=U_{0}(x)-F(t)x,
\end{equation}
where (x,t) is a dimensionless 1+1 space time and
$D=\frac{k_{B}T}{\eta}$. $k_{B}$ is the Boltzmann constant, $T$ is
the temperature and $\eta$ is friction coefficient. The prime stands
for the derivative with respect to the space variable $x$. $j(x,t)$
is the probability current and $\rho(x,t)$ is the particle
concentration. This nonlinear equation yields anomalous diffusion
when $\mu\neq 1$: subdiffusion for $\mu>1$ and superdiffusion for
$0<\mu<1$.
\subsection{Normal diffusion}
\indent For the normal diffusive case $\mu=1$, Eq. (4) recovers the
ordinary Fokker-Planck equation and $j(x,t)$ satisfies
\begin{equation}\label{}
    j(x,t)=-U^{'}(x,t)\rho(x,t)-D\frac{\partial}{\partial
    x}\rho(x,t).
\end{equation}
\indent  The density $\rho(x,t)$ follows a first order non
homogenous linear differential equation, whose formal solution is
\begin{equation}\label{}
    \rho(x,t)=\exp[-\int_{0}^{x}\frac{U^{'}(z,t)}{D}dz]\{c_{0}-\frac{j}{D}\int_{0}^{x}dz\exp[\int_{0}^{z}\frac{U^{'}(y,t)}{D}dy]\}.
\end{equation}
\indent If $F(t)$ changes very slowly with respect to $t$, namely,
its period is longer than any other time scale of the system,
 there exists a quasistatic state. In the steady state, the
 concentration is just a function of space thus the flux becomes a constant $j$.
 Though, unlike typical Brownian motors \cite{6,16,17}, the boundary
conditions are not periodic nor the normalized condition is imposed,
the unknown constant $c_{0}$ and $j$ can be found by imposing the
left reservoir concentration $\rho_{0}\equiv \rho(0)$ and the right
concentration $\rho_{1}\equiv \rho(L)$ as fixed boundary conditions.
We can find that $c_{0}=\rho_{0}$ and
\begin{equation}\label{}
    j(F_{0})=\frac{D\{\rho_{0}-\rho_{1}\exp[-\frac{F_{0}L}{D}]\}}{\int_{0}^{L}\exp[\frac{U_{0}(x)-F_{0}x}{D}]dx}=\frac{D}{I(F_{0})}\{\rho_{0}-\rho_{1}\exp[-\frac{F_{0}L}{D}]\},
\end{equation}
where
\begin{equation}\label{}
    I(F_{0})=\frac{\lambda L D}{Q-F_{0}\lambda L}[\exp(\frac{Q-F_{0}\lambda
    L}{D})-1]+\frac{D(\lambda-1)L}{Q-F_{0}(\lambda-1)L}[\exp(-\frac{F_{0}L}{D})-\exp(\frac{Q-F_{0}\lambda
    L}{D})].
\end{equation}

 \indent The average current is
\begin{equation}\label{}
    J=\frac{1}{\tau}\int^{\tau}_{0}j(F(t))dt=\frac{1}{2}[j(F_{0})+j(-F_{0})].
\end{equation}

\indent For studying the pumping capacity, we consider the situation
in which $J$ tends to zero which corresponds the case in which the
pump in maintaining the maximum concentration difference between the
two reservoirs across the barrier with no net leaking of particle.
This situation is analogous the stalling force in Brownian motors.
From Eqs. (8-10), we can obtain
\begin{equation}\label{}
    \frac{\rho_{1}}{\rho_{0}}=\frac{I(F_{0})+I(-F_{0})}{e^{\frac{F_{0}L}{D}}I(F_{0})+e^{-\frac{F_{0}L}{D}}I(-F_{0})}.
\end{equation}

\subsection{Anomalous diffusion}

\indent When $\mu$ in Eq. (4) is not equal to $1$, we have
\begin{equation}\label{}
    j(x,t)=-U^{'}(x,t)\rho(x,t)-D\frac{\partial}{\partial
    x}\rho^{\mu}(x,t).
\end{equation}
\indent In this case, the explicit expression for $\rho(x,t)$ cannot
be extracted from above equation by using the same method as the
case of $\mu=1$. Here, we use the method presented by Zhao and
coworkers \cite{10} to derive the expressions for $j$ and $\rho$.
Let us assume
\begin{equation}\label{}
    \rho_{m}(x,t)=G(x,t)\rho^{\mu}(x,t), \indent  j_{m}(x,t)=G(x,t)j(x,t),
\end{equation}
where $G(x,t)$ is a factor function. We assume that there exists the
proper $\rho_{m}(x,t)$ and $j_{m}(x,t)$ satisfying diffusion law,
\begin{equation}\label{}
    j_{m}(x,t)=-D\frac{\partial}{\partial x}\rho_{m}(x,t),
\end{equation}
From Eqs. (13) and (14), we have
\begin{equation}\label{}
    j(x,t)=-D\rho^{\mu}(x,t)\frac{\partial}{\partial x}[\ln
    G(x,t)]-D\frac{\partial}{\partial x}\rho^{\mu}(x,t)
\end{equation}
Comparing Eq. (12) with Eq. (15), we have
\begin{equation}\label{}
    G(x,t)=\exp[\frac{\phi_{\mu}(x,t)}{D}],
\end{equation}
where
\begin{equation}\label{}
    \phi_{\mu}(x,t)=\int U^{'}(x,t)\rho^{\mu-1}(x,t)dx.
\end{equation}
\indent It is obvious that $\phi_{\mu}$ represents an effective
$\rho$-dependent potential resulting from the nonlinear diffusive
media. For a quasistatic state, we can get the static solution of
Eq. (4)
\begin{equation}\label{}
    \rho_{s}(x)=[\Psi(x)]_{+}^{\frac{1}{\mu-1}},
\end{equation}
where $[f]_{+}=$max$\{f,0\}$ and

\begin{equation}\label{}
\Psi(x)=\rho_{0}^{{\mu-1}}-\frac{\mu-1}{D\mu}U(x).
\end{equation}
\indent From Eqs. (17-19), we can obtain

\begin{equation}\label{}
    \phi_{\mu}(x,t)=\frac{\mu
    D}{1-\mu}\ln[\Psi(x)].
\end{equation}

So we can rewrite the Eq. (15) as
\begin{equation}\label{}
    j(x,t)=-D\exp[-\frac{\phi_{\mu}(x,t)}{D}]\frac{\partial}{\partial
    x}[\exp(\frac{\phi_{\mu}(x,t)}{D})\rho^{\mu}(x,t)].
\end{equation}
\indent Integrating over $x$ from $0$ to $x$, we have
\begin{equation}\label{}
    \frac{j}{D}\int_{0}^{x}\exp[\frac{\phi_{\mu}(x,t)}{D}]dx=-\int_{0}^{x}\frac{\partial}{\partial
    y}[\exp(\frac{\phi_{\mu}(y,t)}{D})\rho^{\mu}(y,t)]dy.
\end{equation}
\indent So we can get the expression for $\rho^{\mu}(x,t)$
\begin{equation}\label{}
\rho^{\mu}(x,t)=\frac{\exp[\frac{\phi_{\mu}(0,t)}{D}]\rho^{\mu}(0,t)-\frac{j}{D}\int_{0}^{x}\exp[\frac{\phi_{\mu}(x,t)}{D}]dx}{\exp[\frac{\phi_{\mu}(x,t)}{D}]}.
\end{equation}
The constant $j$ can be found by imposing the left reservoir
concentration $\rho_{0}\equiv \rho(0)$ and the right concentration
$\rho_{1}\equiv \rho(L)$ as fixed boundary conditions,
\begin{equation}\label{}
    j(F_{0})=\frac{\theta(F_{0})D}{I(F_{0})}[\rho_{0}^{\mu}-M(F_{0})\rho_{1}^{\mu}],
\end{equation}
where
\begin{equation}\label{}
    M(F_{0})=[\rho_{0}^{\mu-1}+\frac{\mu-1}{D\mu}F_{0}L]^{\frac{\mu}{1-\mu}},
\end{equation}
and
\begin{eqnarray}
\nonumber
  I(F_{0})&=&\int^{L}_{0}[\Psi(x)]^{\frac{\mu}{1-\mu}}dx\\
 &=&\frac{\mu-1}{A}[(\rho_{0}^{\mu-1}-A\lambda
  L)^{\frac{1}{1-\mu}}-\rho_{0}]+\frac{1-\mu}{B}[(C+B L)^{\frac{1}{1-\mu}}-(C+B\lambda
  L)^{\frac{1}{1-\mu}}],
\end{eqnarray}
where $A=\frac{(\mu-1)(Q-F_{0}\lambda L)}{D\mu\lambda}$,
$B=\frac{(\mu-1)[Q+F_{0}(1-\lambda)L]}{D\mu(1-\lambda)L}$ and
$C=\rho_{0}^{\mu-1}-\frac{(\mu-1)Q}{D\mu(1-\lambda)}$.

\indent It must be pointed out that when $\Psi(x)$ is not always
positive there exists a cut off of probability. A cutoff condition
(Tsallis cutoff) yields regions with null probability. This is
because a cutoff of the stationary solution Eq. (18) restricts the
attainable space. Since the probability of particles visiting the
special regions is null, the particles can not pass across pump,
then there is no current. In order to describe this cut off a
function $\theta(F_{0})$ is defined by

\begin{equation}\label{}
 \theta(F_{0})=\left\{
\begin{array}{ll}
   1,& \hbox{$\Psi(x)>0$},  \mbox{for all values of $x$};\\
   0 ,&\hbox{$\mbox{otherwise}$}.\\
\end{array}
\right.
\end{equation}

\indent Similarly, the average current is
\begin{equation}\label{}
    J=\frac{1}{2}[j(F_{0})+j(-F_{0})].
\end{equation}

\indent From Eqs. (24-28), we can obtain the maximum concentration
ratio for $J=0$ and $\rho_{0}=1$,

\begin{equation}\label{}
    \frac{\rho_{1}^{\mu}}{\rho_{0}^{\mu}}=\frac{\theta(F_{0})I(-F_{0})+\theta(-F_{0})I(F_{0})}{\theta(F_{0})I(-F_{0})M(F_{0})+\theta(-F_{0})I(F_{0})M(-F_{0})}.
\end{equation}

\section {Results and discussions}
\indent Our study focus on the current and the maximum concentration
ratio at $J=0$ for normal diffusion, subdiffusion and
superdiffusion. For simplicity, we take $k_{B}=1$, $\eta=1$ and
$L=1$ throughout the study.

\subsection{Normal diffusion $\mu=1$}

\begin{figure}[htbp]
  \begin{center}\includegraphics[width=10cm,height=8cm]{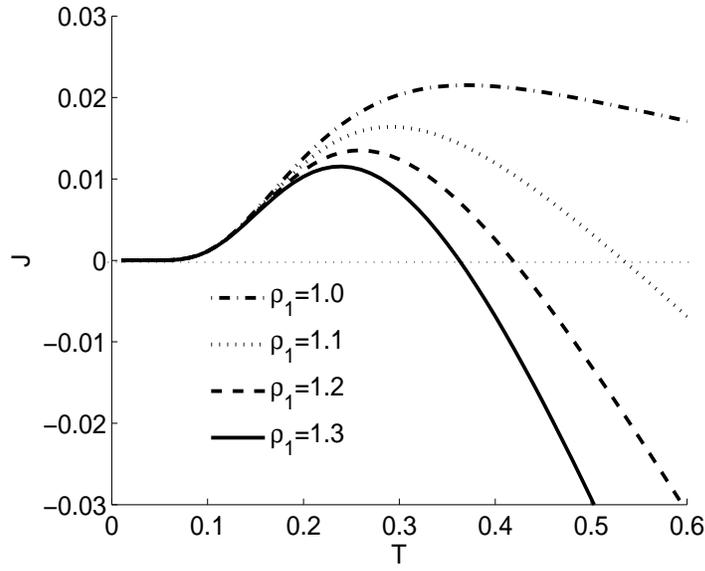}
\caption{Current $J$ versus temperature $T$ for different values of
$\rho_{1}$ at $Q=1$, $\lambda=0.9$, $F_{0}=0.5$, $\rho_{0}=1.0$, and
$\mu=1$. }\label{1}
\end{center}
\end{figure}
\indent In Fig. 2, we present  current as a function of temperature
$T$ for normal diffusion ($\mu=1.0$). For lower values of $\rho_{1}$
the current is larger. When $\rho_{1}$ is more than $\rho_{0}$, the
current is negative for high temperatures. When $T\rightarrow 0$,
the particles cannot pass over the potential barrier and the current
tends to zero. When $T\rightarrow \infty$, the ratchet effect
disappears, the transport is dominated by the concentration
difference and the particles move to the left. Therefore, there
exists an optimized value of $T$ at which the current takes its
positive maximum value. It is note that the similar behavior is also
presented in Ref. 6.

\begin{figure}[htbp]
  \begin{center}\includegraphics[width=10cm,height=8cm]{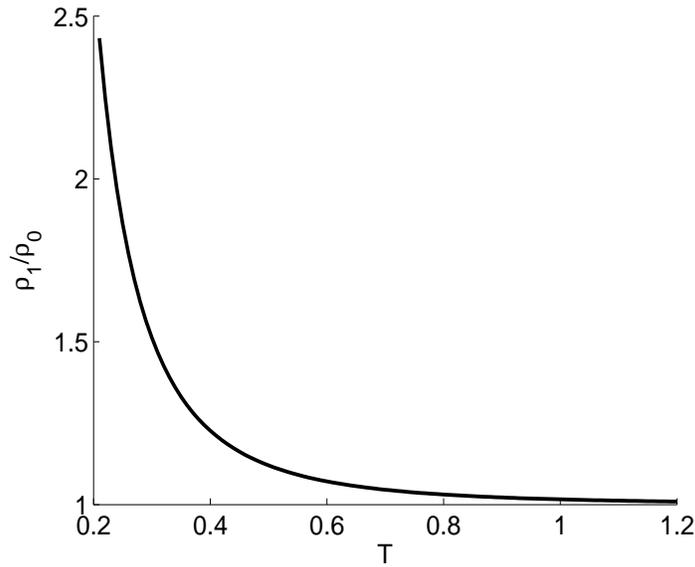}
\caption{Concentration ratio $\rho_{1}/\rho_{0}$  as a function of
temperature $T$ for $J=0$ at $Q=1$, $\lambda=0.9$, $F_{0}=0.5$, and
$\mu=1$. }\label{1}
\end{center}
\end{figure}
\indent Figure 3 shows the ratio $\rho_{1}/\rho_{0}$ as a function
of temperature $T$ for normal diffusion ($\mu=1$). When
$T\rightarrow 0$, no particle can pass over the barrier, thus
$\rho_{1}/\rho_{0}\rightarrow \infty$. As temperature $T$ is
increased, the ratchet effect reduces and the pumping capacity
decreases. Surprisingly, the temperature corresponding to the
maximum current is not the same the temperature at which the
concentration ratio for zero current is maximum. This cause for this
is that zero current induces the maximum concentration ratio, not
the minimum one.

\subsection{Subdiffusion $\mu>1$}
\begin{figure}[htbp]
  \begin{center}\includegraphics[width=10cm,height=8cm]{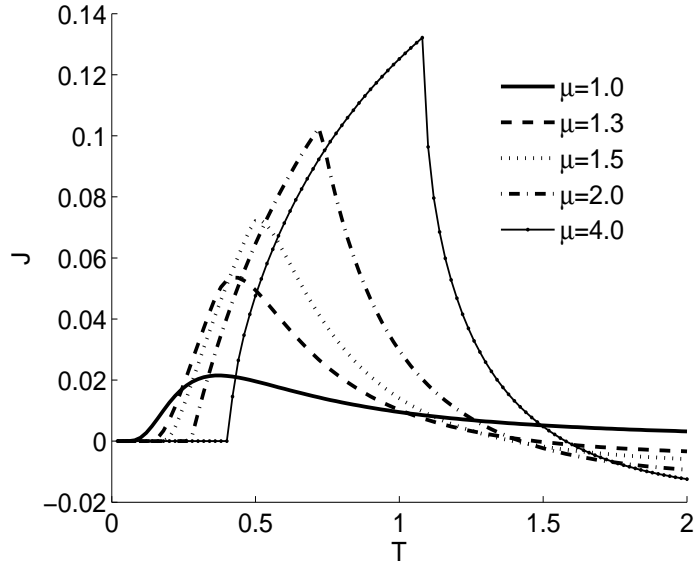}
\caption{Current $J$ as a function of temperature $T$ for different
values of $\mu$ at $Q=1$, $\lambda=0.9$, $F_{0}=0.5$,
$\rho_{0}=1.0$, and $\rho_{1}=1.0$. }\label{1}
\end{center}
\end{figure}

\indent Figure 4 shows the current $J$ versus temperature $T$ for
subdiffusion ($\mu>1$) at $\rho_{1}=\rho_{0}=1.0$. The curve for
normal diffusion is observed to be bell shaped. However, as $\mu$ is
increased, the curve becomes to be not smooth. There exist two
values of temperature at which the curve has inflexions:  the lowest
temperature to obtain a positive current, the optimized temperature
to obtain a maximum current.  For low temperatures, there lies a
finite temperature only above which the particle can pass over the
barrier, otherwise the particles will be confined in both
reservoirs. For high temperatures, the current is negative. However,
when $T\rightarrow \infty$, the current will approach to zero.

\begin{figure}[htbp]
  \begin{center}\includegraphics[width=10cm,height=8cm]{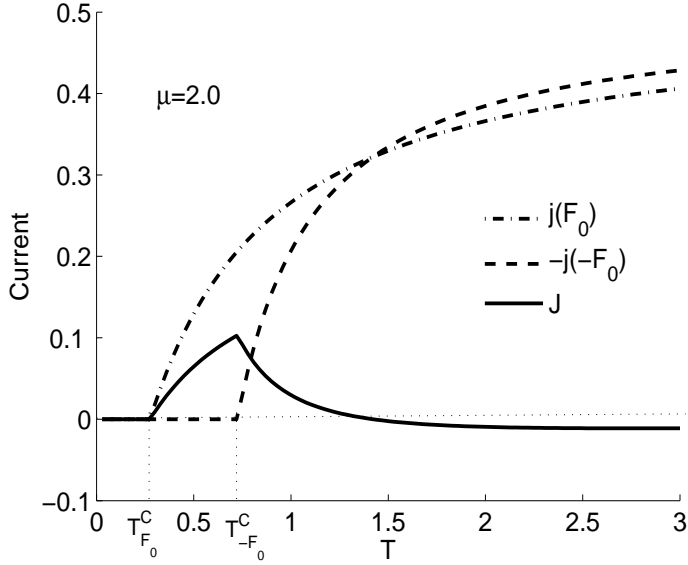}
\caption{Currents $j(F_{0})$, $-j(-F_{0})$ and $J$ versus
temperature $T$ for $\mu=2.0$ at $Q=1$, $\lambda=0.9$, $F_{0}=0.5$,
$\rho_{0}=1.0$, and $\rho_{1}=1.0$. The temperatures for the two
inflexions are $T^{C}_{F_{0}}=0.225$, $T^{C}_{-F_{0}}=0.725$.
}\label{1}
\end{center}
\end{figure}

\indent In order to illustrate the transport behavior for
subdiffusion, the currents $j(F_{0})$, $-j(-F_{0})$ and $J$ as a
function of $T$ for $\mu=2.0$ are shown in Fig. 5. From Eq. (19) and
(27), we can find that there exist two values of temperature

\begin{equation}\label{}
    T^{C}_{\mp F_{0}}=\frac{\mu-1}{\mu}(Q\pm F_{0}\lambda),
\end{equation}
only above which $\Psi (x)$ is always positive, otherwise $\Psi(x)$
may be negative.  Therefore, when $T\leq T^{C}_{\mp F_{0}}$, the
state space becomes disconnected and crossings become forbidden. The
local current $j$ tends to zero. This is because a cut off of the
stationary solution of Eq. (18) restricts the attainable space. From
the figure, we can see that both $j(F_{0})$ and $j(-F_{0})$ are zero
and $J=0$ for $T<T^{C}_{F_{0}}$; $j(F_{0})$ is positive, $j(-F_{0})$
is zero and $J=\frac{1}{2}j(F_{0}$ for
$T^{C}_{F_{0}}<T<T^{C}_{-F_{0}}$; $j(F_{0})$ is positive,$j(-F_{0})$
is negative and $J=\frac{1}{2}[j(F_{0})+j(-F_{0})]$ for
$T>T^{C}_{-F_{0}}$.  The first inflexion is at $T^{C}_{F_{0}}$ and
the second one is at $T^{C}_{-F_{0}}$.

\begin{figure}[htbp]
  \begin{center}\includegraphics[width=10cm,height=8cm]{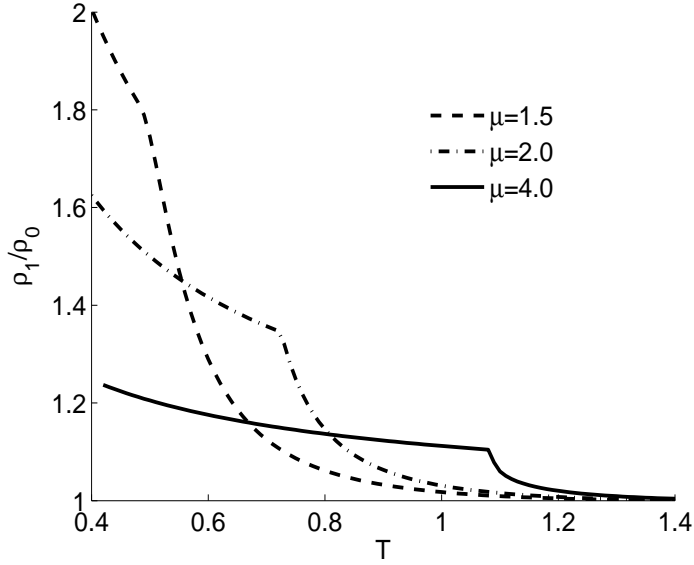}
\caption{Concentration ratio $\rho_{1}/\rho_{0}$ as a function of
temperature $T$ for $J=0$ at  $Q=1$, $\lambda=0.9$, $F_{0}=0.5$,
$\rho_{0}=1.0$, and $\mu=1.5$, $2.0$, $4.0$. }\label{1}
\end{center}
\end{figure}
\indent Figure 6 shows the concentration ratio $\rho_{1}/\rho_{0}$
as a function of $T$ for subdiffusion ($\mu=1.5$, $2.0$, and $4.0$).
The ratio $\rho_{1}/\rho_{0}$ decreases with increasing $T$. The
curve is not smooth and there exists a inflexion  at
$T=T^{C}_{F_{0}}$. For low temperatures, the ratio
$\rho_{1}/\rho_{0}$ decreases with increasing $\mu$, while it
increases with increasing $\mu$ for high temperatures.

\subsection{Superdiffusion $0<\mu<1$}
\begin{figure}[htbp]
  \begin{center}\includegraphics[width=10cm,height=8cm]{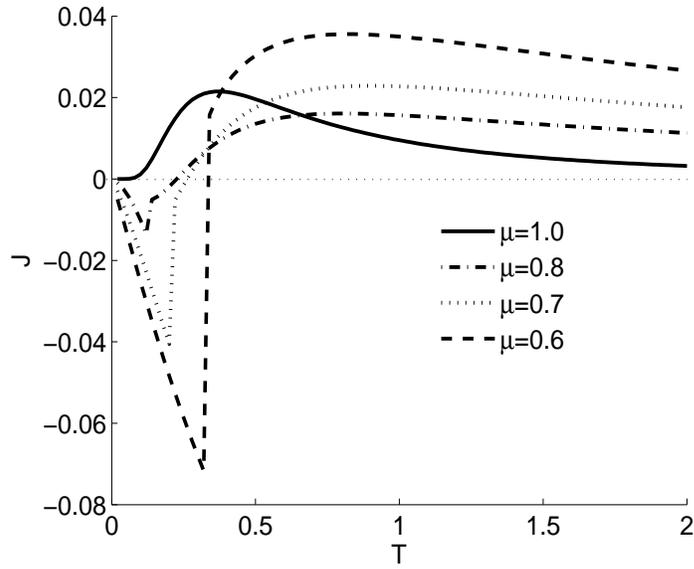}
\caption{Current $J$ as a function of temperature $T$ for different
values of $\mu$ at $Q=1$, $\lambda=0.9$, $F_{0}=0.5$,
$\rho_{0}=1.0$, and $\rho_{1}=1.0$. }\label{1}
\end{center}
\end{figure}
\indent Figure 7 shows current $J$ as a function of temperature $T$
for superdiffusion ($0<\mu<1$) at $\rho_{1}=\rho_{0}$. It is found
that the curve is not smooth and there is a inflexion at which the
negative current takes its maximum value. For low temperatures, the
particles move to the left and the negative current increases with
decreasing $\mu$. As temperature $T$ is increased, the current
becomes to be positive and the positive current increases with
decreasing $\mu$.
\begin{figure}[htbp]
  \begin{center}\includegraphics[width=10cm,height=8cm]{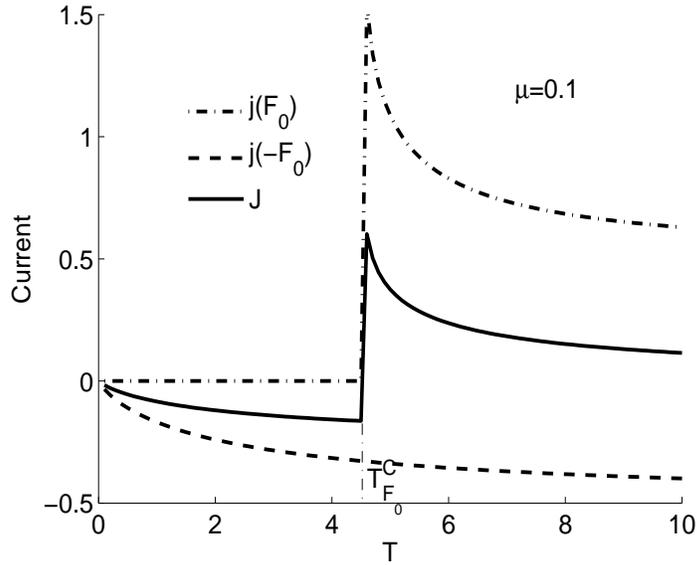}
\caption{Currents $j(F_{0})$, $j(-F_{0})$ and $J$ versus temperature
$T$ for $\mu=0.1$ at $Q=1$, $\lambda=0.9$, $F_{0}=0.5$,
$\rho_{0}=1.0$, and $\rho_{1}=1.0$. The temperature at the inflexion
is $T^{C}_{F_{0}}=4.5$.}\label{1}
\end{center}
\end{figure}

\indent In order to explain the transport behavior for
superdiffusion, the currents $j(F_{0})$, $j(-F_{0})$ and $J$ versus
temperature $T$ for $\mu=0.1$ are shown in Fig. 8. We can find from
Eqs. (19) and (27) that there is a critical value of temperature
\begin{equation}\label{}
    T^{C}_{F_{0}}=-F_{0}\frac{\mu-1}{\mu},
\end{equation}
below which the local current $j(F_{0})$ is zero. When $T$ is less
than $T^{C}_{F_{0}}$, $j(F_{0})$ is zero and $j(-F_{0})$ is
negative, so the current $J=\frac{1}{2}j(-F_{0})$ is negative. When
$T$ is more than $T^{C}_{F_{0}}$, $j(F_{0})$ is more than
$-j(-F_{0})$, so the current $J$ is positive. As temperature $T$ is
increased, $j(F_{0})$ tends to $-j(-F_{0})$, then the current $J$
goes to zero.

\begin{figure}[htbp]
 \begin{center}\includegraphics[width=10cm,height=8cm]{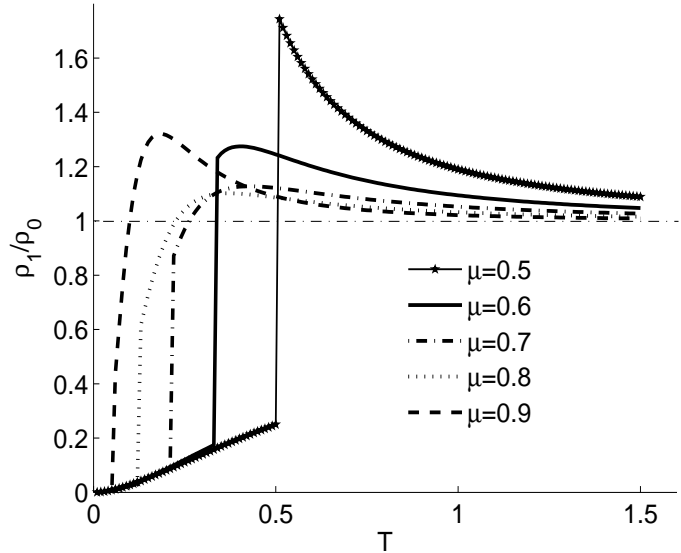}
\caption{Concentration ratio $\rho_{1}/\rho_{0}$ as a function of
temperature $T$ for $J=0$ at  $Q=1$, $\lambda=0.9$, $F_{0}=0.5$,
$\rho_{0}=1.0$, and $\mu=0.5$, $0.6$, $0.7$, $0.8$, $0.9$.
}\label{1}
\end{center}
\end{figure}
\indent Figure 9 shows the concentration ratio $\rho_{1}/\rho_{0}$
as a function of temperature $T$ for superdiffusion ($\mu=0.5$,
$0.6$, $0.7$, $0.8$ and $0.9$). For low temperatures, $T$ is less
than $ T^{C}_{F_{0}}$ and $j(F_{0})$ goes to zero, so the particles
move to the left and $\rho_{1}/\rho_{0}<1$. As $\mu$ is decreased,
the lowest temperature for pumping the particles to the right
increases. Therefore, the pumping system in superdiffusive regime
need a higher temperature than that in normal diffusive regime.

\section{Concluding Remarks}

\indent In this study, we investigate a Brownian pump in nonlinear
diffusive media with an unbiased external force. The pump is
embedded in a finite region and bounded by two particle reservoirs.
The analytical results are obtained in the adiabatic limit. In
normal diffusive regime, current is a peaked function of temperature
and concentration ratio for zero current decreases with increasing
temperature. In subdiffusive regime, current is forbidden for low
temperatures and negative for high temperatures. The concentration
ratio for zero current decreases with increasing temperature, also.
In superdiffusive regime, the transport compared with normal
diffusion exhibits an opposite direction for low temperatures. There
exists a finite temperature at which the concentration ratio takes
its maximum value. For anomalous diffusion, there exist inflexions
in the curves for $J$ vs $T$ and $\rho_{1}/\rho_{0}$. This is
because a cut off of the stationary solution Eq. (18) restricts the
attainable space.

\begin{center}
    \textbf{{ACKNOWLEDGMENTS}}
\end{center}
 \indent The work was supported by the National
Natural Science Foundation of China under Grant No. 30600122 and
GuangDong Provincial Natural Science Foundation under Grant No.
06025073.

\end{document}